\documentclass[showpacs,aps,prd,nofootinbib,floatfix,amsmath,amssymb]{revtex4}
\usepackage{graphicx}
\begin{document}


\title{CP-odd static electromagnetic properties  of the $W$ gauge boson and the $t$ quark via the anomalous $tbW$ coupling}
\author{J. Hern\'andez-S\' anchez}
\affiliation{Instituto de
Ciencias B\' asicas e Ingenier\'\i a, Universidad Aut\' onoma del
Estado de Hidalgo, Carretera Pachuca-Tulancingo Km. 5.5, C. P.
42814, Pachuca, Hgo., M\' exico.}
\author{C. G. Honorato}
\author{F. Procopio}
\author{G. Tavares-Velasco}
\author{J. J. Toscano}
\affiliation{Facultad de
Ciencias F\'{\i}sico Matem\'aticas, Benem\'erita Universidad
Aut\'onoma de Puebla, Apartado Postal 1152, Puebla, Pue., M\' exico}

\date{\today}

\begin{abstract}
In the framework of the electroweak chiral Lagrangian, the one-loop
induced effects of the anomalous $tbW$ coupling, which includes both
left- and right-handed complex components, on the static
electromagnetic properties of the $W$ boson and the $t$ quark are
studied. The attention is focused mainly on the CP-violating
electromagnetic properties. It is found that the $tbW$ anomalous
coupling  can induce both CP-violating moments of the $W$ boson,
namely, its electric dipole ($\widetilde{\mu}_W$) and magnetic
quadrupole ($\widetilde{Q}_W$) moments. As far as the $t$ quark is
concerned, a potentially large electric dipole moment $(d_t)$ can
arise due to the anomalous $tbW$ coupling. The most recent bounds on
the left- and right-handed parameters from $B$ meson physics lead to
the following estimates $\widetilde{\mu}_W\sim 4\times
10^{-23}-4\times 10^{-22}$ e$\cdot$ cm and $\widetilde{Q}_W\sim
10^{-38}-10^{-37}$ e$\cdot$ cm$^2$, which are $7$ and $14$ orders of
magnitude larger than the standard model (SM) predictions, whereas
$d_t$ may be as large as $10^{-22}$ e$\cdot$cm, which is about $8$
orders of magnitude larger than its SM counterpart.

\end{abstract}


\maketitle

\section{Introduction}

The only source of CP violation in the standard model (SM) is the
Cabbibo-Kobayashi-Maskawa (CKM) phase, which seems to be the origin
of CP violation in nondiagonal processes \cite{CKMND}, as suggested
by the experimental data on $B$-$\Bar{B}$ mixing \cite{BBM}. Several
studies \cite{DS} suggest, however, that the CKM phase has a rather
marginal effect on CP-violating flavor-diagonal processes such as
the electric dipole moments (EDM) of elementary particles, which
means that this class of properties may be highly sensitive to any
new physics effects. While the static electromagnetic properties of
the leptons have been long studied both theoretically and
experimentally, those of the heaviest SM particles still require
more attention. In particular, the CP-violating properties of the
$W$ boson and the top quark are extremely suppressed in the SM (the
$W$ boson EDM arises up to two loops and the top quark EDM arises
first at three loops), so their study may shed light on the origin
of CP violation. It is thus worth analyzing alternative sources of
CP violation that may manifest themselves via the static
electromagnetic properties of the $W$ boson and the top quark.

The fact that the on-shell $WW\gamma$ vertex is gauge-independent
has long motivated the study of its behavior under radiative
corrections because this allows one to estimate the sensitivity of
the Yang-Mills sector to any physics beyond the Fermi scale. Much
theoretical work has gone into studying the CP-even static
electromagnetic properties of the $W$ boson, but its experimental
determination still awaits a higher experimental precision. As far
as the CP-odd electromagnetic properties are concerned, they are by
far more suppressed within the SM, thereby offering an ideal
laboratory for searching for any new physics effects. By invoking
Lorentz and electromagnetic gauge invariance, regardless of $C$,
$P$, or $T$ conservation, the on-shell $WW\gamma$ vertex can be
characterized by five independent form factors
\cite{FF}.\footnote{This is a general result, which is true for any
no self-conjugate vector field even if it is electrically neutral,
in which case the monopolar coupling cannot exist and so $g_1$
vanishes at any order of perturbation theory \cite{NPTT,TT1}.} Three
of these form factors are CP-even $(g_1,\kappa, \Delta Q)$ and two
are CP-odd, $(\widetilde{\kappa},\widetilde{Q})$. While $g_1$ and
$\kappa$ are already generated at the level of the classical action,
$\Delta Q$ first arises at the one-loop level. The anomalous
contributions to $\kappa$ and $\Delta Q$  have been calculated in
the SM \cite{WWgSM} and several of its extensions \cite{WWgNP}. As
for the CP-odd form factors, they are naturally suppressed because
they can only arise at the one-loop level or higher orders in any
renormalizable theory. Despite their suppression, the scrutiny of
the CP-violating $W$ boson properties may provide relevant
information for our knowledge of CP violation. The electric dipole
and magnetic quadrupole moments can be generated at the one-loop
level via the $\widetilde{\kappa}$ form factor in some extended
models. Since the presence of a trace involving the $\gamma_5$ Dirac
matrix is necessary in order to generate a Levi-Civitta tensor, it
is clear that these CP-violating moments can only be induced at the
one-loop level via a fermionic loop. It has been shown that this
class of effects can arise in theories including both left- and
right-handed fermion currents with a complex phase
\cite{Burgess,TT1}. This possibility has already been explored
within the context of left-right symmetric models \cite{Burgess},
though the respective contribution was found highly suppressed due
to the experimental constraints on the $W_L-W_R$ mixing.

As far as the static electromagnetic properties of the top quark are
concerned, the forthcoming years will see a vigorous boost in the
theoretical interest and experimental scrutiny of this particle's
properties. Specifically, the largest priority at the CERN large
hadron collider (LHC) is the study of the top quark fundamental
properties, and further studies are planned at the next linear
collider (NLC) via top quark pair production. Interesting
experimental and theoretical prospects are open due to the fact that
the top quark has a mass of the order of the Fermi scale, which
poses the question whether it is just an ordinary quark or a
composite particle. The top quark decays very quickly, mainly into a
$bW$ pair, before any hadronization takes place, which may allow one
to examine its properties without the presence of any unwanted QCD
effects, which invariably would swamp the processes involving the
light quarks. This peculiarity opens the door to the scrutiny of the
top quark electromagnetic properties, thereby allowing the
possibility of detecting a nonvanishing CP-violating electric dipole
moment (EDM). Several studies have been devoted to analyze the
electric or weak dipole form factors of the top quark and the so
induced CP violation. Along these lines, several studies on CP
violation in $\bar{t}t$ production have been pursued in the context
of hadron \cite{HC}, $e^+e^-$ \cite{EPC}, and $\gamma \gamma$
\cite{PPC} colliders. Despite its suppression in the SM, the top
quark EDM can be significantly enhanced in a broad class of
beyond-the-SM extensions. For instance, studies within the context
of multi-Higgs models \cite{EDMMH} have shown that the top quark EDM
may be several orders of magnitude larger than the SM prediction.

Since the study of the CP-odd static electromagnetic properties of
the $W$ boson and the top quark may hint to the origin of CP
violation, it is worth considering any possible sources of this
class of effects. In this work, we are interested in the potential
CP-violating effects of the most general dimension-four $tbW$
coupling, which involves both left- and right-handed components with
a complex phase, on the CP-odd static electromagnetic properties of
the $W$ boson and the top quark. The relevance of this coupling is
evident from the fact that the top quark decays mainly into a $bW$
pair. Although there is a plenty of theoretical work that has
considered its tree-level effects, it is also interesting to study
its possible impact in one-loop processes. Due to the large mass of
the top quark, which is the only known quark with a mass of the
order of the electroweak scale, it has been conjectured that it may
induce new dynamics effects and even more play a special role in the
mechanism of mass generation, which has attracted the interest on
the study of any anomalous contributions to its couplings to other
SM particles. Even more, the copious production of top quark events
expected at the LHC will allow us to study more carefully the top
quark properties and examine possible new physics effects induced by
this particle. Very interestingly, this class of new physics effects
would arise at the electroweak scale, in contrast with other type of
new physics efffects that are expected to arise at heavier energy
scales.

The role that the $tbW$ coupling might play in a scenario in which
the mass is not generated via the Higgs mechanism has been examined
by several authors \cite{tbW} through diverse phenomenological
studies \cite{EWCL,Peccei:1989kr}. Instead of considering a specific
model, we will adopt a model-independent approach by considering the
$tbW$ anomalous contributions in the context of an electroweak
chiral Lagrangian (EWCL) \cite{EWCL} in which the $SU_L(2)\times
U_Y(1)$ symmetry is nonlinearly realized as it is assumed that the
Higgs boson is very heavy or does not exist at all. We can think of
this scenario as the one in which the EWCL parametrizes unknown
physics that is not dictated by the Higgs mechanism. Several authors
have already studied this vertex in this context, and diverse
scenarios have been taken into account to obtain limits on the left-
and right-handed couplings
\cite{LIMPHASES1,LIMPHASES2,LIMEWO,LIMUNITARITY}. Along these lines,
some top quark production mechanisms have been considered
\cite{PRODUCTION}. In particular, some hadronic processes have been
used to impose limits on the CP-violating phases associated with the
general left- and right-handed structure of the $tbW$ coupling
\cite{LIMPHASES1,LIMPHASES2}. These results will be used below to
estimate the values of the $W$ electric dipole and magnetic
quadrupole moments. Thus, the main aim of this work is using the
most recent bounds on the anomalous terms of the $tbW$ vertex to
predict the effects of this coupling on the static electromagnetic
properties of the $W$ boson and the top quark. This approach is in
accordance with the spirit of the effective Lagrangian approach.

This paper has been organized as follows. The most general structure
of the $tbW$ vertex is introduced in the context of the EWCL in Sec.
II, whereas the respective contribution to the CP-violating static
electromagnetic properties of the $W$ boson and the top quark are
calculated in Sec. III. Section IV is devoted to analyze our
results, and the conclusions are presented in Sec. V.

\section{Theoretical framework}
 If the Higgs mechanism is not realized in nature, whatever
is the true mechanism responsible for the electroweak symmetry
breaking, it would open unexpected avenues for new physics effects.
In particular, new sources of CP violation might show up. The
unknown new physics can be parametrized using the effective
Lagrangian technique in which the electroweak symmetry is
nonlinearly realized. The resultant Lagrangian is known as the EWCL.
In this approach, the Higgs doublet is replaced by a dimensionless
matrix field that transforms nonlinearly under the $SU_L(2)\times
U_Y(1)$ group \cite{tbW,EWCL}:
\begin{equation}
\Sigma =\exp\Big(\frac{i\phi^a\sigma^a}{v}\Big),
\end{equation}
where $\phi^a$ ($a=1,2,3$) are Goldstone bosons, $\sigma^a$ are the
Pauli matrices, and $v$ is the Fermi scale. Under the $SU_L(2)\times
U_Y(1)$ group, $\Sigma$ transform as:
\begin{equation}
\Sigma'=L\Sigma R^\dag,
\end{equation}
where
\begin{eqnarray}
L&=&\exp\Big(\frac{i\alpha^a \sigma^a}{2}\Big), \\
R&=&\exp\Big(\frac{i\beta \sigma^3}{2}\Big),
\end{eqnarray}
with $\alpha^a$ and $\beta$ being the parameters of the $SU_L(2)$
and $U_Y(1)$ groups, respectively. From these expressions, it is
easy to see that the Goldstone bosons fields transform nonlinearly
under the electroweak group. In this scheme, the gauge fields are
defined in the following way
\begin{eqnarray}
\hat{W}_\mu&=&\frac{\sigma^aW^a_\mu}{2i}, \\
\hat{B}_\mu&=&\frac{\sigma^3B_\mu}{2i}.
\end{eqnarray}

To define the most general expression for the charged current, it is
necessary to introduce some bosonic and fermionic Lorentz
structures. We need the following Lorentz tensors:
\begin{eqnarray}
\Sigma^a_\mu&=&-\frac{i}{2}Tr[\sigma^a \Sigma^\dag D_\mu \Sigma ],
\\
\Sigma^a_{\mu \nu}&=&-iTr[\sigma^a \Sigma^\dag
[D_\mu,D_\nu]\Sigma],
\end{eqnarray}
where
\begin{equation}
D_\mu \Sigma=\partial_\mu \Sigma-g\hat{W}_\mu \Sigma +g'\Sigma
\hat{B}_\mu.
\end{equation}
\noindent The charged fields are given by the following relations
\begin{eqnarray}
\Sigma^\pm_\mu&=&\frac{1}{\sqrt{2}}(\Sigma^1_\mu \mp
i\Sigma^2_\mu), \\
\Sigma^\pm_{\mu \nu}&=&\frac{1}{\sqrt{2}}(\Sigma^1_{\mu \nu} \mp
i\Sigma^2_{\mu \nu}).
\end{eqnarray}
In the unitary gauge ($\phi^a=0$), these expressions become
$\Sigma^\pm_\mu=g/2\,W^\pm_\mu$ and
\begin{equation}
\Sigma^\pm_{\mu \nu}=g[W^\pm_{\mu \nu}\pm ie(W^\pm_\mu
A_\nu-W^\pm_\nu A_\mu)\pm igc_W(W^\pm_\mu Z_\nu-W^\pm_\nu Z_\mu)],
\end{equation}
where $c_W$ stands for $\cos\theta_W$. We also need the following
fermion operators:
\begin{eqnarray}
\Delta_L&=&\bar{t}P_Lb, \ \ \  \Delta^\mu_L=\bar{t}\gamma^\mu
P_Lb, \\
\Delta_R&=&\bar{t}P_Rb, \ \ \  \Delta^\mu_R=\bar{t}\gamma^\mu
P_Rb,
\end{eqnarray}

\begin{eqnarray}
\Delta^{\mu \nu}_L&=&\bar{t}\sigma^{\mu \nu}P_Lb, \ \ \ \bar{\Delta}^\mu_L=i\bar{t}P_L\bar{D}^\mu b, \\
\Delta^{\mu \nu}_R&=&\bar{t}\sigma^{\mu \nu}P_Rb, \ \ \
\bar{\Delta}^\mu_R=i\bar{t}P_R\bar{D}^\mu b,
\end{eqnarray}
where $\bar{D}_\mu=\partial_\mu -ieQA_\mu$ is the electromagnetic
covariant derivative, $\sigma_{\mu
\nu}=(i/2)[\gamma_\mu,\gamma_\nu]$, and $P_L(P_R)$ is the
left-handed(right-handed) projector.

Using the above expressions, the most general Lagrangian for the
charged currents can be written as
\begin{eqnarray}
{\cal
L}^{CC}&=&\sqrt{2}a_L\Delta^\mu_L\Sigma^+_\mu+\sqrt{2}a_R\Delta^\mu_R\Sigma^+_\mu+\frac{1}{\Lambda}\Big[ib_L\Delta_L\bar{D}^\mu
\Sigma^+_\mu+ib_R\Delta_R\bar{D}^\mu \Sigma^+_\mu \nonumber \\
&&+c_L\bar{\Delta}^\mu_L\Sigma^+_\mu+c_R\bar{\Delta}^\mu_R\Sigma^+_\mu+d_L\Delta^{\mu
\nu}_L\Sigma^+_{\mu \nu}+d_R\Delta^{\mu \nu}_R\Sigma^+_{\mu
\nu}\Big]+{\rm H.c.},
\end{eqnarray}
with $\Lambda$ being an energy scale. We have introduced the
$\sqrt{2}$ factor in order to recover the SM value ($g/\sqrt{2}$)
for the left-handed coupling in the appropriate limit.\footnote{We
do not introduce operators proportional to $\bar{D}^\mu\bar{t}$
because they are not independent. In fact, after integration by
parts one obtains $\bar{D}^\mu \bar{t}b
\Sigma^+_\mu=-\bar{t}\bar{D}^\mu b \Sigma^+_\mu -\bar{t}b\bar{D}^\mu
\Sigma^+_\mu$.} Although this Lagrangian induces the vertices $tbW$,
$tbW\gamma$ and $tbWZ$, we only show the first one:
\begin{eqnarray}
\label{lagtbW} {\cal L}_{tbW}&=&
\frac{g}{2}\Big\{\sqrt{2}\bar{t}(a_LP_L+a_RP_R)b W^+_\mu
+\frac{1}{\Lambda}\Big[i\bar{t}(b_LP_L+b_RP_R)b\partial^\mu
W^+_\mu \nonumber \\
&&+i\bar{t}(c_LP_L+c_RP_R)\partial^\mu bW^+_\mu+ 2\bar{t}\sigma^{\mu
\nu}(d_LP_L+d_RP_R)b W^+_{\mu \nu}\Big]\Big\}+{\rm H.c.}
\end{eqnarray}
We will only consider the renormalizable part of this coupling as it
is expected to give the dominant contribution to the $WW\gamma$ and
$tt\gamma$ vertices. Therefore, all terms proportional to
$1/\Lambda$ will be neglected from now on.

A word of caution is in order here. Although we have presented an
specific theoretical framework for the presence of an anomalous
$tbW$ coupling, our calculation below remains valid for any class of
theory predicting an interaction as the one given in Eq.
(\ref{lagtbW}). In this context, our results are meant to examine
the impact of the anomalous $tbW$ coupling on the static properties
of the $W$ boson and the top quark, but they are not mean to test
the EWCL scenario. This is beyond the reach of the present work.

\section{CP-odd static electromagnetic properties of the $W$ boson and the top quark}

We now present the explicit calculation of the anomalous $tbW$
coupling to the CP-odd static electromagnetic properties of the $W$
boson and the top quark.

\subsection{The $W$ boson EDM and MQM}

We now turn to discuss the structure of the $WW\gamma$ vertex and
the contribution from the anomalous $tbW$ coupling. The most general
on-shell $W_\alpha (p-q)W_\beta (-p-q) A_\mu (2q)$ vertex can be
written as

\begin{eqnarray}
\label{elecprop} \Gamma_{\alpha \beta
\mu}&=&i\,e\bigg(g_1\left[2\,p^\mu g^{\alpha \beta}+4\,(q_\beta
g_{\alpha \mu}-q_\alpha g_{\beta \mu})\right] +2\Delta \kappa
(q_\beta g_{\alpha \mu}-q_\alpha g_{\beta \mu})+\frac{4\,\Delta
Q}{m^2_{W}}\,p_\mu q_\alpha q_\beta\nonumber\\&+&2
\widetilde{\kappa}\,\epsilon_{\alpha \beta \mu
\lambda}q^\lambda+\frac{4\,\widetilde{Q}}{m^2_{W}}\,q_\beta
\epsilon_{\alpha \mu \lambda \rho}p^\lambda q^\rho\bigg),
\end{eqnarray}
where all the momenta are incoming. In a renormalizable theory, the
form factors $\Delta \kappa$, $\Delta Q$, $\widetilde{\kappa}$, and
$\widetilde{Q}$ always arise via radiative corrections. The magnetic
(electric) dipole moment $\mu_{W}$ ($\widetilde{\mu}_W$) and the
electric (magnetic) quadrupole moment $Q_{W}$ ($\widetilde{Q}_{W}$)
are given in terms of the electromagnetic form factors as follows
\begin{eqnarray}
\label{mu_W}
\mu_{W}&=&\frac{e}{2\,m_{W}}(2+\Delta \kappa), \\
\label{Q_W}
Q_{W}&=&-\frac{e}{m^2_{W}}(1+\Delta \kappa+\Delta Q),\\
\label{mutilde}
\widetilde{\mu}_W&=&\frac{e}{2\,m_{W}}\widetilde{\kappa},\\
\label{Qtilde} \widetilde{Q}_{W}&=&-\frac{e}{m^2_{W}}(
\widetilde{\kappa}+\widetilde{Q}).
\end{eqnarray}

\begin{figure}
\centering
\includegraphics[width=2.5in]{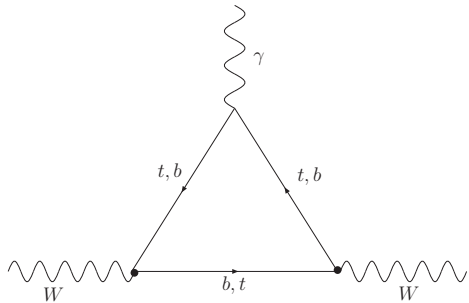}
\caption{\label{WWg}Feynman diagrams contributing to the on-shell
$WW\gamma$ vertex. The dot denotes an anomalous $tbW$ vertex. }
 \end{figure}

The contribution of the $tbW$ coupling arises from the triangle
diagrams shown in Fig. \ref{WWg}. We will concentrate only on the
CP-odd contribution, although there are also contributions to the
CP-even form factors. As already mentioned, at this order of
perturbation theory, the renormalizable part of the $tbW$ coupling
can only contribute to one CP-odd form factor, namely,
$\widetilde{\kappa}$, which reads

\begin{equation}
\widetilde{\kappa}=\frac{\alpha}{8s_W^2\pi^2}\left(Q_t
F(x_t,x_b)+Q_b F(x_b,x_t)\right) Im\left(a_L a_R^\dagger\right),
\end{equation}
where $x_a=m_a/m_W$ and
\begin{equation}
F(x,y)=4\,x\,y\left(\log\left(\frac{x}{y}\right)+\left(1-x^2+y^2\right)\frac{f(x,y)}{2\chi(x,y)}\right),
\end{equation}
with
\begin{equation}
f(x,y)=\log\left(\frac{1-(x^2+y^2)-\delta (x,y)}{1-(x^2+y^2)+\delta
(x,y)}\right),
\end{equation}
and $\chi^2(x,y)=(x^2 + y^2 - 1)^2 - 4 x^2y^2$. This result is free
of ultraviolet divergences, which is a consequence of the fact that
only the renormalizable part of the $tbW$ vertex has been
considered.

\subsection{The top quark EDM}

The most general $tt\gamma$ vertex function is given by

\begin{equation}
\Gamma_\mu=i\,\bar{t}(p_1)\left(\gamma^\mu
F_t+i\sigma^{\mu\nu}q_\nu\left(a_t+\gamma^5 d_t\right)\right)t(p_2),
\end{equation}
where $F_t$ is the electric charge of the top quark, $a_t$ is its
MDM, and $d_t$ is its EDM. At tree level $Ft=2/3 e$, whereas $a_t$
and $d_t$ arise via radiative corrections.

\begin{figure}
\centering
\includegraphics[width=4in]{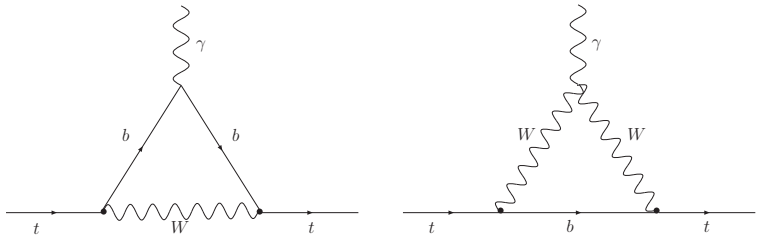}
\caption{\label{ttg}Feynman diagrams contributing to the on--shell
$\bar{t}t\gamma$ vertex.}
 \end{figure}

The anomalous $tbW$ coupling of Eq. (\ref{lagtbW})  induces an EDM
for the top quark through the Feynman diagrams shown in Fig.
\ref{ttg}, where all the particles are taken on-shell. After some
calculation via the unitary gauge, one can extract the coefficient
of the $i \gamma^5\sigma_{\mu\nu}q^\nu$ term from the
$\bar{t}t\gamma$ vertex function. Here $q_\mu$ is the photon four
momentum. This leads to

\begin{equation}
\label{topEDM} d_t=\frac{N_c\,\alpha}{32
\pi}\,\frac{m_b}{m_W}\frac{e}{m_W}\,\left(Q_W F_W(x_b,x_W)+
Q_b\,F_b(x_b,x_W)\right)\,Im\left(a_L a_R^*\right),
\end{equation}

\noindent with $x_i=m_i/m_t$, $N_c=3$, $Q_b=-1/3$, and $Q_W=-1$. The
$F_W$ and $F_b$ functions stand for the contribution of the Feynman
diagram where the photon emerges from the $W$ boson and the $b$
quark line, respectively. They are given by

\begin{equation}
F_W(x_b,x_W)=\left(x_b^2-4x_W^2-1\right)f_1(x_b,x_W)-\left(x_b^4+4
x_W^4-5x_b^2x_W^2-3x_W^2-2x_b^2+1\right)f_2(x_b,x_W),
\end{equation}

\begin{equation}
F_b(x_b,x_W)=\left(x_b^2-4x_W^2-1\right)f_1(x_W,x_b)+\left(x_b^4+4
x_W^4-5x_b^2x_W^2-3x_W^2-2x_b^2+1\right)f_2(x_W,x_b),
\end{equation}
\noindent with
\begin{eqnarray}
f_1(x,y)&=&2+\left(1-y^2+x^2\right)\log\left(\frac{y}{x}\right)+\sqrt{(1-x^2-y^2)^2-4x^2y^2}\,{\rm
sech}^{-1}\left(\frac{2 x y}{x^2+y^2-1}\right),\\
\nonumber
f_2(x,y)&=&-\log\left(\frac{y}{x}\right)-\frac{1+x^2-y^2}{\sqrt{(1-x^2-y^2)^2-4x^2y^2}}\,{\rm
sech}^{-1}\left(\frac{2 x y}{x^2+y^2-1} \right).
\end{eqnarray}
For the same reason argued when discussing the $W$ boson EDM, the
calculated top quark EDM is free of ultraviolet divergences.

\section{Results and discussion}
\begin{figure}
\centering
\includegraphics[width=4in]{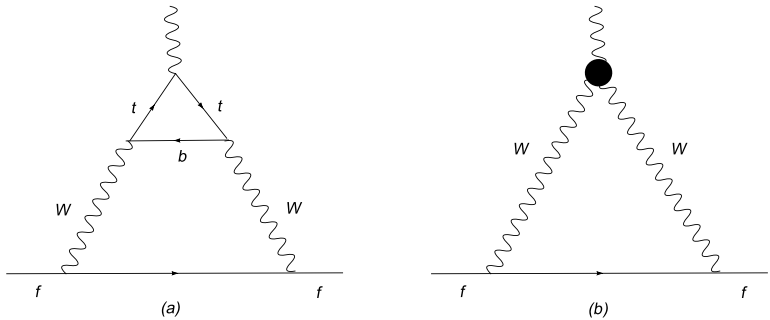}
\caption{\label{twoloop}Feynman diagrams contributing to the EDM of
a fermion via the $W$ EDM when (a)the fermions circulating in the
$WW\gamma$ loop have a mass of the same order of magnitude than the
$W$ boson mass and (b)the internal fermions are much heavier than
the $W$ boson, in which case an effective $WW\gamma$ vertex can be
used.}
 \end{figure}

We turn to discuss the numerical results. First of all, we will
discuss the most recent bounds on the anomalous $tbW$ coupling from
$B$ meson physics. Secondly, we will consider these bounds to
predict the order of magnitude expected for the $W$ boson and the
top quark EDMs. We would like to emphasize that our main purpose is
to obtain a prediction for the $tbW$ effects on the CP-violating
static electromagnetic properties of the $W$ boson and the top
quark, rather than using our calculation to obtain a bound for the
anomalous $tbW$ coupling, for which we could use the neutron EDM for
instance. However, a careful study of the effects of the one-loop
induced $W$ boson EDM on the  nucleon EDM would require a two-loop
calculation, which is beyond the purpose of this work. It is worth
to discuss this point with more extent. In Ref. \cite{MQ}, an
effective vertex parametrizing the EDM of the $W$ boson was used to
calculate a one-loop induced fermion EDM. Somewhat erroneously, we
may want follow the same approach here and use the calculation of
Ref. \cite{MQ} to obtain a bound on the $tbW$ anomalous coupling.
This could be done safely if the fermions circulating in the loop
were much heavier than the external $W$ boson, in which case the
one-loop $WW\gamma$ vertex of Fig. \ref{WWg} could be parametrized
as an effective tree-level vertex an inserted into the one-loop
$\bar f f \gamma$ vertex to calculate the fermion EDM [see Fig.
\ref{twoloop} (b)]. Such an approximation is not valid when the
$WW\gamma$ loop includes fermions with a mass of the same order of
magnitude than the $W$ boson mass, as occurs with the contribution
of the $tbW$ vertex. In such a case, the two-loop diagram of Fig.
\ref{twoloop} (a) must be evaluated.

Although there are prospects for the direct measurement of the $tbW$
coupling at the LHC and the planned future colliders, it has been
pointed out in Ref. \cite{CP} that CLEO data on $b\to s\gamma$ are
already more constraining on the right-handed $tbW$ coupling than
what would be achievable at any planned future collider. Thus, the
bounds discussed below will be very useful to assess the impact of
this vertex on the $WW\gamma$ and $tt\gamma$ vertices.

\subsection{Bounds on the anomalous $tbW$ coupling}

It is customary to parametrize the left- and right-handed parameters
of Eq. (\ref{lagtbW}) in the following way
\begin{eqnarray}
a_L&=&1+\kappa_Le^{i\phi_L}, \\
a_R&=&\kappa_Re^{i\phi_R},
\end{eqnarray}
with $\kappa_{L,R}$ and $\phi_{L,R}$ real parameters. It follows
that
\begin{equation}
Im(a_La^\dag_R)=-\kappa_R \sin\phi_R+\kappa_L \kappa_R
\sin(\phi_L-\phi_R).
\end{equation}
In the above expressions, the SM left-handed coupling was explicitly
introduced along with a deviation characterized by the $\kappa_L$
and $\phi_L$ parameters. In order to make predictions, we need to
assume some values for these parameters. For this purpose, we will
consider the bounds reported in the literature, such as the ones
obtained in Ref. \cite{LIMPHASES1} from $B$ decay processes:
\begin{eqnarray}
\label{Bdecay}
\kappa_L \sin\phi_L&<&3\times 10^{-2}, \\
\kappa_R \sin\phi_R&<&10^{-3}.
\end{eqnarray}
There are also limits on the right-handed parameters derived from
the CLEO Collaboration data on the $b\to s\gamma$ decay
\cite{LIMPHASES2}:
\begin{eqnarray}
\label{CLEO}
\kappa_R \cos\phi_R&<&4\times 10^{-3}, \\
\kappa_R \sin\phi_R&<&10^{-3}.
\end{eqnarray}
In addition, current data on CP-conserving process allows $\kappa_L$
to be as large as $0.2$ \cite{CP,LIMEWO,LIMUNITARITY}. As far as the
$\kappa_R$ parameter is concerned, it seems to be more suppressed
than the corresponding left-handed one, as suggested by Eq.
(\ref{CLEO}) and also from the result obtained in Ref.
\cite{Yamada}, where it was found that $-5\times
10^{-2}<\kappa_R<10^{-2}$.

In the following, we will estimate the EDM of the $W$ boson and the
top quark using the following values: $\kappa_L\sin\phi_L<3\times
10^{-2}$, $\kappa_R \sin\phi_R<10^{-3}$, and $\kappa_R<4\times
10^{-3}$.

\section{The EDM of the $W$ boson}
In the context of renormalizable theories with the simultaneous
presence of left- and right-handed fermion currents, the electric
dipole and magnetic quadrupole moments are proportional at the
one-loop level:
\begin{equation}
\widetilde{Q}_W=-\Big(\frac{2}{m_W}\Big)\widetilde{\mu}_W.
\end{equation}
This means that  $\widetilde{Q}_W$ is suppressed with respect to
$\widetilde{\mu}_W$ by a factor of the order of $10^{-16}$, provided
that units of e and cm are used. However, this hierarchy might not
hold at higher orders. Using the known values for the SM parameters,
the electric dipole and magnetic quadrupole moments of the $W$ boson
can be written as
\begin{eqnarray}
\widetilde{\mu}_W&=&-4\times 10^{-19} Im(a_La^\dag _R) \quad {\rm e}\cdot {\rm cm}, \\
\widetilde{Q}_W&=&1.98\times 10^{-34}Im(a_La^\dag _R) \quad {\rm
e}\cdot {\rm cm}^2.
\end{eqnarray}

The constraints of Eqs. (\ref{Bdecay}) and (\ref{CLEO}) pose two
scenarios of interest, for which we get an estimate for the $W$
electric dipole and magnetic quadrupole moments:
\begin{itemize}
\item {SM-like $a_L$ and complex $a_R$:}
\begin{eqnarray}
\widetilde{\mu}_W&=&4\times 10^{-19}\kappa_R \sin\phi_R  \ {\rm e}\cdot {\rm cm}<4\times 10^{-22} \ {\rm e}\cdot {\rm cm}, \\
\widetilde{Q}_W&=&-1.98\times 10^{-34} \kappa_R \sin\phi_R \ {\rm
e}\cdot {\rm cm}^2<-1.98\times 10^{-37} \ {\rm e}\cdot {\rm cm}^2.
\end{eqnarray}

\item {Complex $a_L$ and purely real $a_R$:}
\begin{eqnarray}
\widetilde{\mu}_W&=&-4\times 10^{-19} \kappa_R\kappa_L\sin\phi_L  \ {\rm e}\cdot {\rm cm}<-4.8\times 10^{-23} \ {\rm e}\cdot {\rm cm}, \\
\widetilde{Q}_W&=&1.98\times 10^{-34}\kappa_R\kappa_L\sin\phi_L \
{\rm e}\cdot {\rm cm}^2<2.38\times 10^{-38} \ {\rm e}\cdot {\rm
cm}^2.
\end{eqnarray}
\end{itemize}
In the most general scenario, both $a_L$ and $a_R$ are complex, but
the values for $\widetilde{\mu}_W$ and $\widetilde{Q}_W$ are similar
to those obtained in the first scenario above.

It is worth comparing our results with those previously reported in
the literature. Of course the standard to which all results should
be compared with is the SM prediction. As already mentioned, in the
SM $\widetilde{Q}_W$ first arises at the two-loop level, whereas
$\widetilde{\mu}_W$ appears up to three-loop order. It has been
estimated that $\widetilde{\mu}_W$ and $\widetilde{Q}_W$ are of the
order of $10^{-29}$ e$\cdot$ cm  \cite{SMED,SMED2} and $-10^{-51}$
e$\cdot$ cm$^2$ \cite{SMMQ}, respectively. Beyond the SM, most of
the studies have focused on $\widetilde{\mu}_W$, with the exception
of Ref. \cite{TT2}, in which both $\widetilde{\mu}_W$ and
$\widetilde{Q}_W$ were estimated. In sharp contrast with the
negligibly small SM predictions, some of its extensions predict
values several orders of magnitude larger. For instance, a value of
$10^{-22}$  e$\cdot$cm for $\widetilde{\mu}_W$ was estimated in
left-right symmetric models \cite{SMED,Burgess}, and a similar
result was found in supersymmetric models, which induce this moment
via one-loop diagrams mediated by charginos and neutralinos
\cite{SMED,SSM}. The $\widetilde{\mu}_W$ moment has also been
estimated in multi-Higgs models, in which it can be induced  at the
two-loop level. For instance it was found that $\widetilde{\mu}_W
\sim 10^{-20}-10^{-21}$ e$\cdot$ cm in the two-Higgs doublet model
\cite{THDM}. A similar value was found in the context of the
so-called 331 models, whose Higgs sector also induces
$\widetilde{\mu}_W$ at the two-loop level \cite{331}. From these
results, we can conclude that our estimate for $\widetilde{\mu}_W$
in the context of the EWCL lies within the range of the predictions
obtained from other renormalizable SM extensions.

It is interesting to mention that the CP-odd static electromagnetic
properties of the $W$ boson have been estimated in Ref. \cite{TT2}
within the context of the linear electroweak effective Lagrangian
(LEWEL), which do assume the existence of the Higgs boson, in
contrast with the EWCL. It was shown that the most general $HWW$
coupling, which includes both CP-even and CP-odd components, can
give rise to both CP-odd $WW\gamma$ form factors $\tilde \kappa$ and
$\tilde Q$. The numerical estimates obtained for appropriate values
of the unknown parameters are $\widetilde{\mu}_W \sim 3-6\times
10^{-20}$ e$\cdot$ cm and $\widetilde{Q}_W\sim -10^{-36}$ e$\cdot$
cm$^2$ \cite{TT2}. We thus can conclude that the values for
$\widetilde{\mu}_W$ and $\widetilde{Q}_W$ induced by the anomalous
$tbW$ coupling, which are about 7 and 17 orders of magnitude above
their respective SM predictions, are one order of magnitude smaller
that those generated by an anomalous $HWW$ coupling.

\subsection{The top quark EDM}

Once Eq. (\ref{topEDM}) is numerically evaluated, one obtains
\begin{eqnarray}
d_t&=&(3.08-5.73)\times 10^{-19}\,Im\left(a_L a_R^*\right)\, {\rm
e\cdot cm}, \nonumber
\\
&=&-2.65\times 10^{-19}\,Im\left(a_L a_R^*\right)  \, {\rm e\cdot
cm} ,
\end{eqnarray}
where the positive (negative) contribution corresponds to the the
Feynman diagram where the photon emerges from the boson (quark)
line. It should be mentioned that $d_t$ develops an imaginary part,
which is almost twice larger than the real one. The appearance of an
imaginary (absortive) part is not usual in the static
electromagnetic properties of light particles, but in this case it
arises as a consequence of the fact that, being the top quark so
heavy, $m_t>m_W+m_b$.

Bearing in mind the constraints of Eqs. (\ref{Bdecay}) and
(\ref{CLEO}), we will explore the two scenarios discussed above. We
obtain in the first scenario

\begin{equation}
|d_t|\lesssim 2.65\times 10^{-22} \ {\rm e\cdot cm},
\end{equation}
whereas the latter scenario leads to
\begin{equation}
|d_t|\lesssim 7.95\times 10^{-23}\ {\rm e\cdot cm}.
\end{equation}
In this case, the constraint $\kappa_R<10^{-2}$ was used.

We also would like to compare our results with those obtained within
the framework of other theories. As already mentioned, in the SM the
top quark EDM arises first at three loops and it has been estimated
to be of the order of $10^{-30}$ e$\cdot$cm \cite{SMED,SMTOPEDM}.
Beyond the SM, the top quark EDM has also received some attention.
For instance, in multi--Higgs models, values for $d_t$ lying in the
range $10^{-20}-10^{-21}$ e$\cdot$cm have been estimated
\cite{EDMMH}. We can thus conclude that our prediction, which is
compatible with the constraints imposed by $B$ meson physics, is
about eight orders of magnitude larger than the SM one but one or
two orders of magnitude smaller than that obtained in multi--Higgs
models.

\section{Final remarks}
Up to now, the  mechanism responsible for electroweak symmetry
breaking remains the most puzzling piece of the SM. The Higgs
mechanism, although satisfactory to generate the masses of the
theory, might be a mere mathematical artifact lacking of any
connection with the physical reality. It is therefore important to
be open-minded to any potential scenario that may give rise to
sizeable new physics effects. Due to its heavy mass, it has been
conjectured that the top quark may play a special role in the mass
generation. Along these lines, it is worth considering potential new
physics effects induced by the top quark. In this paper, we have
explored the possible CP-violating effects arising from the most
general dimension-four $tbW$ coupling. In particular, the impact of
such a coupling on the CP-odd electromagnetic properties of the $W$
boson and the top quark was studied. Analytical expressions were
obtained, and their numerical values were estimated by considering
the most recent bounds on the $tbW$ coupling phases from $B$ meson
physics, as reported in the literature.

It was found that the most promising scenario corresponds to an
anomalous $tbW$ coupling with a SM-like left-handed part and a
complex right-handed component. In such a scenario, the resultant
predictions for the $W$ CP-odd static electromagnetic properties are
$\widetilde{\mu}_W < 4\times 10^{-22}$ e$\cdot$ cm and
$\widetilde{Q}_W<1.98 \times 10^{-37}$ e$\cdot$ cm$^2$, which are 7
and 14 orders of magnitude larger than the respective SM prediction,
whereas the top quark EDM $d_t$ may be up to eight orders of
magnitude larger than the SM contribution. Another interesting
scenario corresponds to an anomalous $tbW$ coupling with a complex
left-handed component and a purely real right-handed component, but
in such a case the static properties of the $W$ boson and the top
quark are one order of magnitude smaller than in the precious
scenario. We have also compared our results with those obtained in
the framework of other beyond-the-SM theories. It was found that our
results are about one order of magnitude smaller than those
predicted by other theories such as multi-Higgs doublet models.
Since an anomalous $tbW$ coupling can arise in a scenario where the
Higgs boson is absent, it is interesting to compare our results with
those obtained in theories where the Higgs boson is the one
responsible for the electroweak symmetry breaking. For instance, in
Ref. \cite{TT2} a general $HWW$ coupling involving both CP-even and
CP-odd components was considered, which gives rise to both
$\widetilde\mu_W$ and $\widetilde Q_W$. Our results, obtained in the
Higgsless EWCL scenario, are about one order of magnitude smaller
than those induced in the scenario where at least one relatively
light Higgs boson is present.

The EDM of the $W$ boson can contribute significantly to the EDM of
light fermions, such as the electron and the neutron, which are
under frequent experimental scrutiny. This fact was exploited by the
authors of Ref. \cite{MQ}, in which the experimental upper bound on
the neutron electric dipole moment, $d_n<10^{-25}$ e$\cdot$ cm, was
used to obtain the upper bound $\widetilde{\mu}_W<10^{-20}$ e$\cdot$
cm. Our result for $\widetilde{\mu}_W$ is consistent with this
result. On the other hand, the direct measurement of the CP-odd
structure of the $WW\gamma$ vertex might be in the range of
sensitivity of next linear colliders (NLC) or CLIC \cite{NLC}, which
will operate as a $W$ factory. The possibility of extracting some
CP-odd asymmetries from these colliders has been examined by several
authors, mainly in a model-independent manner via the effective
Lagrangian technique \cite{AS}.

Alternatively, we could use our calculation to obtain bounds on the
anomalous $tbW$ coupling from the EDM of the neutron. Although in
Ref. \cite{MQ} an effective $WW\gamma$ vertex was used to calculate
the respective contribution to the neutron EDM and the resultant
constraint on the CP-odd $WW\gamma$ coupling can be used to
constrain the anomalous $tbW$ coupling, a precise estimate would
require a two-loop calculation, which is beyond the purpose of this
work.

\acknowledgments{We acknowledge support from Conacyt (M\' exico)
under grants 50764 and J50027-F.}

\end{document}